# СРАВНЕНИЕ СРЕДНИХ ПО ВРЕМЕНИ ПОТЕРЬ МОЩНОСТИ В БИПОЛЯРНЫХ ТРАНЗИСТОРАХ С ИЗОЛИРОВАННЫМ ЗАТВОРОМ И КОМБИНИРОВАННЫХ СИТ-МОП-ТИРИСТОРАХ

*А. С.Кюрегян*[1], *А. В. Горбатюк*[2], *Б. В. Иванов*[3],

[1]Всероссийский электротехнический институт им. В.И. Ленина.
[2]Физико-технический институт им. А.Ф. Иоффе Российской академии наук.
[3]Санкт-Петербургский государственный электротехнический университет «ЛЭТИ» им. В.И. Ульянова (Ленина).

**Аннотация**

Проведено двумерное численное моделирование процессов переключения эквивалентных кремниевых биполярных транзисторов с изолированным затвором типа CSTBT и комбинированных СИТ-МОП-тиристоров (КСМТ) из блокирующего состояния в проводящее и обратно. Показано, что и при включении, и при выключении энергия коммутационных потерь в КСМТ больше, чем в полностью эквивалентном CSTBT. Поэтому средняя по времени мощность $\overline{P}$, рассеиваемая в КСМТ, становится меньше, чем в эквивалентном CSTBT, только при большой длительности $T_{ON}$ импульса тока. Однако уменьшение времени жизни $\tau_0$ неравновесных носителей заряда в СИТ позволяет существенно снизить коммутационные потери КСМТ, сохраняя его преимущество в открытом состоянии. Вследствие этого для каждого набора параметров CSTBT можно подобрать такое $\tau_0$ в «почти эквивалентном» КСМТ, что мощность $\overline{P}$, рассеиваемая в КСМТ, будет меньше, чем в эквивалентном CSTBT, в любом заданном диапазоне значений амплитуды $J_a$ и длительности $T_{ON}$ импульсов тока.

## 1. Введение

В настоящее время биполярные транзисторы с изолированным затвором (БТИЗ) являются наиболее эффективными и распространенным ключевыми элементами для преобразователей средней мощности [1,2]. Наилучшими характеристиками обладают БТИЗ траншейной конструкции, содержащие дополнительный стоп-слой между $p^+$-коллектором и высокоомной *n*-базой [3] (в англоязычной литературе такой прибор обычно называется Carrier Storage Trench Bipolar Transistor, или CSTBT), однако для его производства необходима технологическая база очень высокого уровня. Гораздо более прост в изготовлении комбинированный СИТ-МОП-тиристор (КСМТ), который является точным функциональным аналогом CSTBT. Это гибридный прибор, содержащий высоковольтный тиристор с электростатическим управлением (СИТ) и управляющий низковольтный МОП-транзистор, расположенные на двух кристаллах и соединенные по каскодной схеме [4]. Экспериментальные исследования опытных образцов КСМТ [5,6] показали, что эти приборы как будто не уступают

CSTBT. Однако сравнение проводилось между приборами, которые могли отличаться не только принципиально разными способами управления инжекцией электронов, но и рядом геометрических и электрофизических параметров чипов. Эти особенности экспериментов [5,6] исключают возможность выявления фундаментальных физических отличий, которые могут быть найдены только путем сравнения характеристик приборов различных типов, которые эквивалентны в том смысле, что все параметры их высоковольтных частей идентичны.

Недавно в работе [7] такое сравнение было проведено путем численного моделирования статических характеристик «эквивалентных» кремниевых CSTBT и КСМТ. Было показано, что блокирующая способность в закрытом состоянии практически одинакова, а в открытом состоянии падение напряжения $U_{ON}$ на CSTBT больше, чем на КСМТ, при плотности тока анода $J_a > 10$ А/см$^2$. Однако этой информации недостаточно для полного сравнения, так средняя мощность $\overline{P}$, рассеиваемая коммутаторами в процессе работы, включает в себя и коммутационные потери. Если период следования $T^{-1}$ и длительность $T_{ON}$ прямоугольных импульсов тока много больше характерных времен выключении $t_{off}$ и включения $t_{on}$, то величину $\overline{P}$ можно представить в виде $\overline{P} = T^{-1}\left(P_{ON}T_{ON} + W_c\right)$, где $P_{ON} = U_{ON}J_{ON}$ - установившаяся мощность, рассеиваемая в открытом состоянии, $W_c = W_{on} + W_{off}$ - полная энергия коммутационных потерь,

$$W_{on} = \int_0^{T_{ON}} \left[P(t) - P_{ON}\right]dt, \quad W_{off} = \int_{T_{ON}}^{T} P(t)\,dt, \qquad (1)$$

$P(t) = U_a(t)J_a(t)$ - мгновенная рассеиваемая мощность, причем энергии $W_{on}$ и $W_{off}$ не зависят от $T$ и $T_{ON}$. Целью настоящей работы, продолжающей исследование [7], является расчет переходных характеристик $U_a(t)$, $J_a(t)$ и сравнительный анализ средних потерь мощности $\overline{P}$ в «эквивалентных» кремниевых CSTBT и КСМТ.

## 2. Конструкции приборов и метод моделирования их характеристик.

Объектами исследования были CSTBT и КСМТ, конструкция которых схематично изображена на Рис. 1. Геометрические и электрофизические параметры исходных эпитаксиальных структур и диффузионных слоев приведены в таблицах 1-3 из работы [7], однако в некоторых случаях использовались иные значения параметров $\tau_0 = \tau_{n0} = \tau_{p0}$, входящих в формулу Шокли−Рида для скорости рекомбинации через глубокий уровень. Расчеты переходных и необходимых для полноты анализа статических характеристик проводились, как и в [7], с помощью программы TCAD SENTAURUS фирмы SYNOPSYS [8]. Использованные в настоящей работе метод построения сетки конечных элементов и зависимости кинетических и рекомбинационных коэффициентов от уровня легирования также описаны в [7]. Дополнительно были учтены зависимости подвижностей электронов и дырок от напряженности электрического поля, которая изменяется от единиц до сотен тысяч В/см в процессе коммутации.

В использованной нами модели коммутация приборов осуществлялась путем мгновенного изменения потенциалов затворов управляющих МОП-транзисторов с -20 В до +20 В (при включении) и обратно (при выключении). Согласно расчетам [7] при этом приведенное сопротивление $R_{\mathrm{dsON}}$ открытого МОП-транзистора в CSTBT равно примерно 1 мОм см$^2$. Это

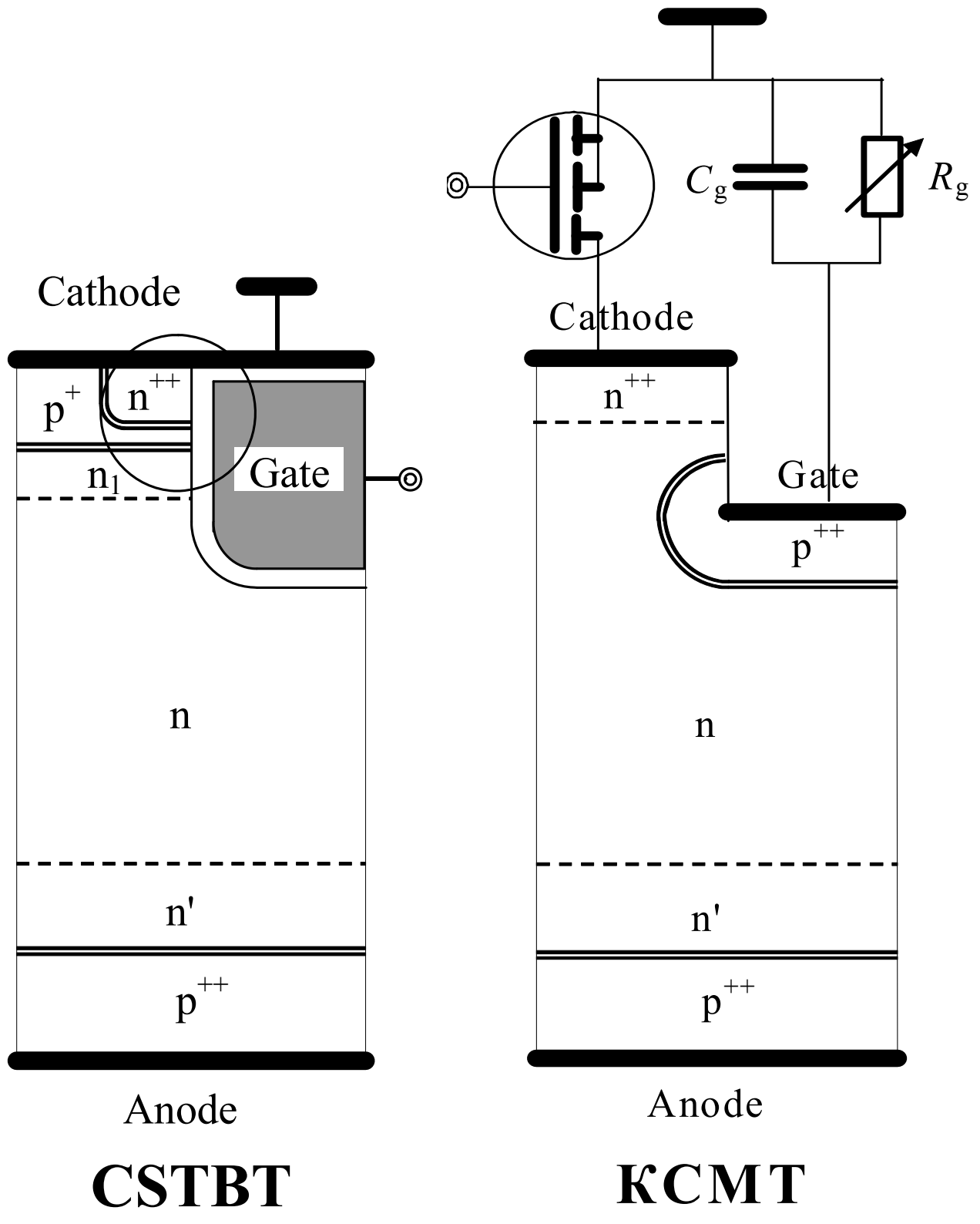

Рис. 1. Схематичное изображение поперечной ячейки CSTBT и комбинированного СИТ-МОП-тиристора (КСМТ). Окружностью выделен интегрированный в основной чип CSTBT полевой транзистор, который в КСМТ изготовляется на отдельном чипе.

же значение $R_{\mathrm{dsON}}$ мы использовали для управляющего МОП-транзистора в КСМТ. Вместо низковольтного стабилитрона [9], стабистора [4] или регулирующего МОП-транзистора [10,11], соединяющего затвор СИТ с заземленным электродом КСМТ, мы использовали в качестве регулирующего элемента КСМТ параллельную RC-цепочку, емкость $C_g$ которой варьировалась от 0 до 20 нФ/см$^2$, а сопротивление $R_g$ мгновенно изменялась от $R_{g0} = 1$ мОм см$^2$ (при включении) до $R_{g1} = (10\text{-}10^7)$ Ом см$^2$ (при выключении) и обратно. Эти упрощения позволили, во-первых, сократить очень значительное время моделирования переходных процессов в КСМТ, во-вторых, фактически исключить влияние параметров драйверов на резуль-таты сравнения характеристик двух приборов и, в-третьих, выяснить, как параметры регулирующих элементов в цепи затвора СИТ влияют на его переключение.

## 3. Результаты моделирования и их обсуждение.

Как уже было отмечено во Введении, в открытом состоянии падение напряжения на КСМТ заметно меньше, чем на эквивалентном CSTBT, но в качественном отношении вольт-амперные характеристики совпадают. Это обстоятельство проиллюстрировано на Рис. 2 для одного из наборов параметров. Переходные характеристики этих приборов различаются не только количественно, но и качественно: включение КСМТ, в отличие от CSTBT, происходит с заметной задержкой $t_d$ (см. Рис. 3). Задержка включения CSTBT отсутствует в нашей модели, так как мы считали, что время заряда входной емкости затвор-катод равно нулю[1]. Однако для начала включения КСМТ необходимо еще, чтобы напряжение на барьерной емкости обратно смещенного перехода затвор-катод $C_{cg} \sim 1$ нФ/см$^2$ уменьшилось до напряжения отсечки. Расчеты показывают, что $t_d$ может изменяться от десятков наносекунд до десятков микросекунд в зависимости от параметров СИТ и регулирующего элемента КСМТ, а также от режима включения. Анализ этих зависимостей выходит за рамки настоящей работы и ему будет посвящена специальная статья. Для наших целей достаточно отметить, что энергия $W_{on}$ потерь при включении КСМТ (и, тем более, полная энергия $W_c$, которая обычно много больше $W_{on}$) практически не зависит от $C_g$, $R_g$ и параметров затвора СИТ. Поэтому большинство расчетов в настоящей работе было выполнено при значениях $C_g = 0$, $R_g = 10$ МОм·см$^2$.

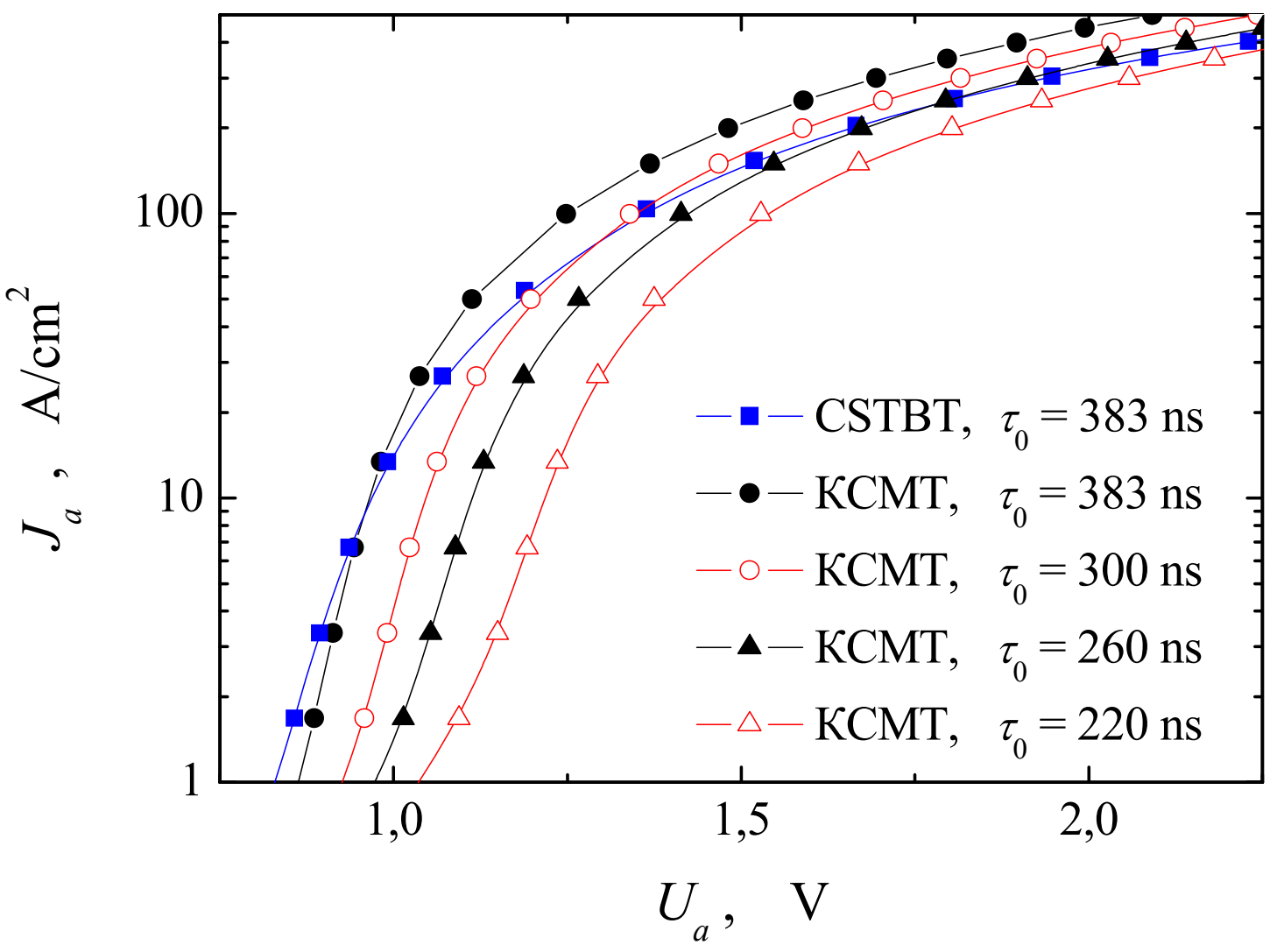

Рис. 2. Вольт-амперные характеристики КСМТ с $x_g = 3$ мкм (кружки) и CSTBT (квадраты) при $U_b = 1.4$ кВ и различных значениях параметров $\tau_0$.

[1] На практике оно обычно равно 20-50 нс.

Количественные различия между коммутационными характеристикам состоят в том, что и при включении, и при выключении потери в КСМТ больше, чем в эквивалентном CSTBT. Для значений параметров, соответствующих рисунку 3, энергии потерь приведены в Таблице. Причина этого, очевидно, состоит в различии зарядов $Q_{пр}$ неравновесных электронов и дырок в высокоомных базах: в CSTBT $Q_{пр}$ меньше из-за относительно низкой эффективности инжекции электронов катодным эмиттером, поэтому работа, необходимая для введения этого заряда при включении и его извлечения при выключении также оказывается меньше, чем в СИТ. Таким образом, одна и та же причина увеличивает статические и уменьшает коммутационные потери в CSTBT.

| Параметр | Обозначение и ед. измерения | КСМТ | CSTBT |
|---|---|---|---|
| Мощность, рассеиваемая прибором в открытом состоянии | $P_{ON}$ , кВт/см$^2$ | 124,5 | 135,3 |
| Энергия потерь при включении | $W_{ON}$ , мДж/см$^2$ | 1,47 | 0,402 |
| Энергия потерь при выключении | $W_{OFF}$ , мДж/см$^2$ | 10,2 | 6,86 |
| Суммарные коммутационные потери | $W_c$ , мДж/см$^2$ | 11,67 | 7,262 |

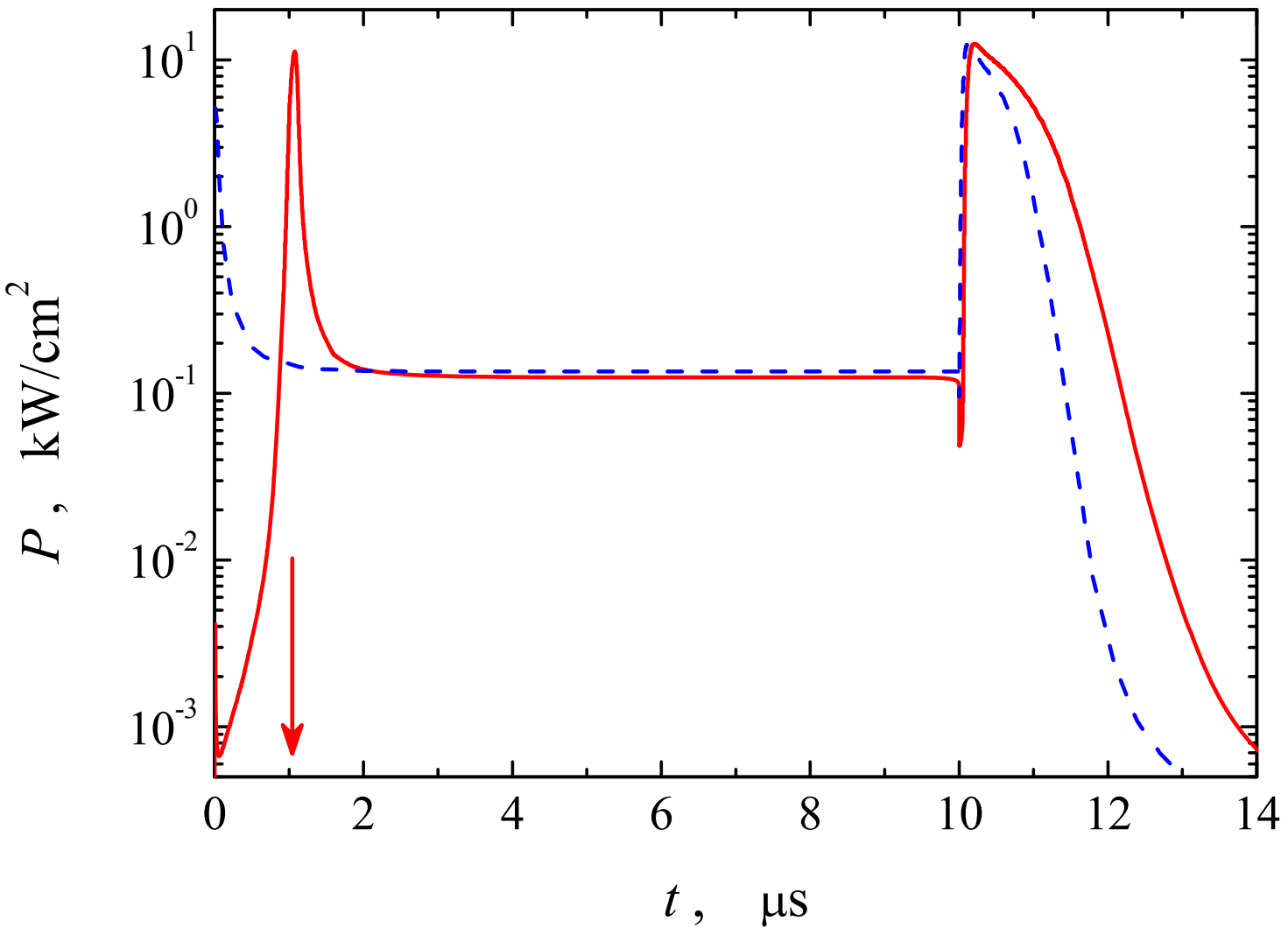

Рис. 3. Зависимости мощности, рассеиваемой КСМТ (сплошная линия) и CSTBT (штриховая линия) с $U_b$ = 1,4 кВ, $\tau_0 = 383$ нс, $R_g = 10^7$ Ом·см$^2$ и $C_g = 0$ при пропускании прямоугольного импульса тока с длительностью 10 мкс и амплитудой 100 А/см$^2$ через сопротивление нагрузки 5 Ом. Стрелкой указано время $t_d$ задержки включения КСМТ.

Для наглядного сравнения эффективности двух типов коммутаторов удобно использовать зависимость отношения $\eta(T_{ON}) = \overline{P}_{KCMT} / \overline{P}_{CSTBT}$ средних рассеиваемых ими мощностей от длительности импульса тока. Пример таких зависимостей приведен на Рис. 4. Как видно, в этом конкретном случае КСМТ эффективней эквивалентного CSTBT только при $T_{ON} > 100$ мкс. Однако результаты работы [7] указывают на то, что обычно рабочая плотность тока $J_{us} = (20 \div 100)$ А/см$^2$ заметно больше величины $J_= = (0.1 \div 10)$ А/см$^2$ (см. Рис. 2 и 6), при которой мощности $P_{ON}$, рассеиваемые в открытом состоянии эквивалентными КСМТ и CSTBT совпадают. Это означает, что можно уменьшать время жизни $\tau_0$ в СИТ, сохраняя его преимущество в открытом состоянии и уменьшая (или даже исключая) повышенные коммутационные потери. Результаты расчетов, приведенные на Рис. 4, показывают, что потери в КСМТ со сниженным всего на 22 % временем жизни $\tau_0$ меньше, чем в «прежнем» CSTBT, при всех значениях $T_{ON}$. Это преимущество не очень велико, но при снижении $\tau_0$ в КСМТ на 43 % оказывается, что «прежний» CSTBT рассеивает мощность уже на 43 % большую, чем «модифицированный» КСМТ, в наиболее актуальном диапазоне значений $T_{ON} \le 10$ мкс.

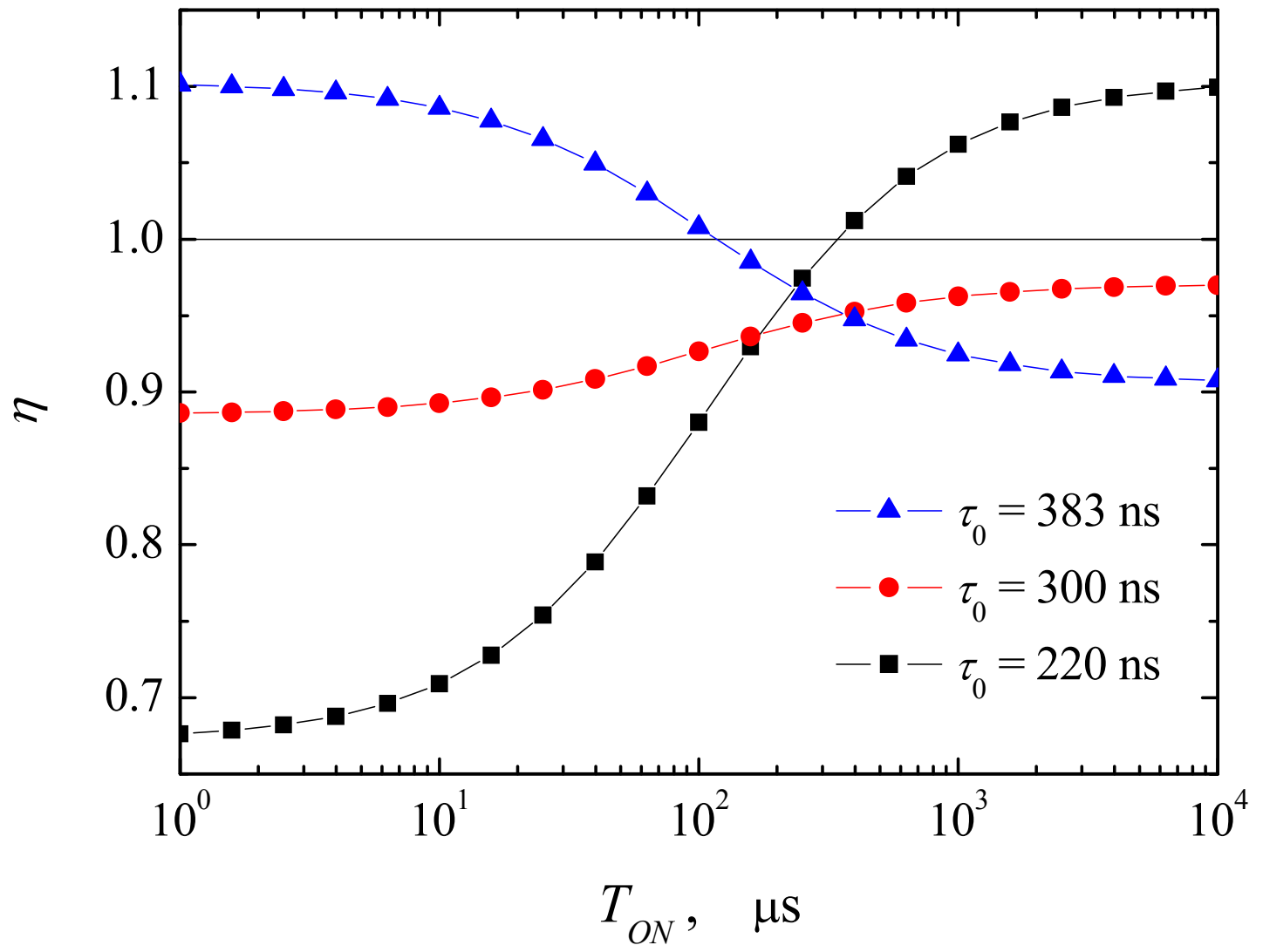

Рис. 4. Зависимость отношения $\eta$ средних мощностей, рассеиваемых КСМТ и CSTBT от длительности импульса тока с амплитудой 157 А, при $U_b = 1.4$ кВ, $R = 5$ Ом, $\tau_0 = 383$ в CSTBT и различных $\tau_0$ в КСМТ.

С ростом плотности тока статические потери в CSTBT растут быстрее, чем в КСМТ (см. Рис. 2), а коммутационные – медленнее (при постоянном сопротивлении нагрузки). Соответствующим образом видоизменяются зависимости $\eta(T_{ON})$, как это изображено на Рис. 5.

Следует подчеркнуть, что неравенство $J_{us} >> J_=$ выполняется при всех значениях параметров эквивалентных КСМТ и CSTBT [7]. Поэтому описанный выше результат является весьма общим, то есть для каждого набора параметров CSTBT можно подобрать такое время жизни в «почти эквивалентном» КСМТ, что величина $\eta(T_{ON})$ будет заметно меньше 1 в заданном диапазоне значений амплитуды $J_a$ и длительности $T_{ON}$ импульсов тока. В качестве еще одного примера, иллюстрирующего это утверждение, на Рис. 6 приведены зависимости $\eta(T_{ON})$ для КСМТ и CSTBT с напряжением пробоя $U_b = 4.6$ кВ.

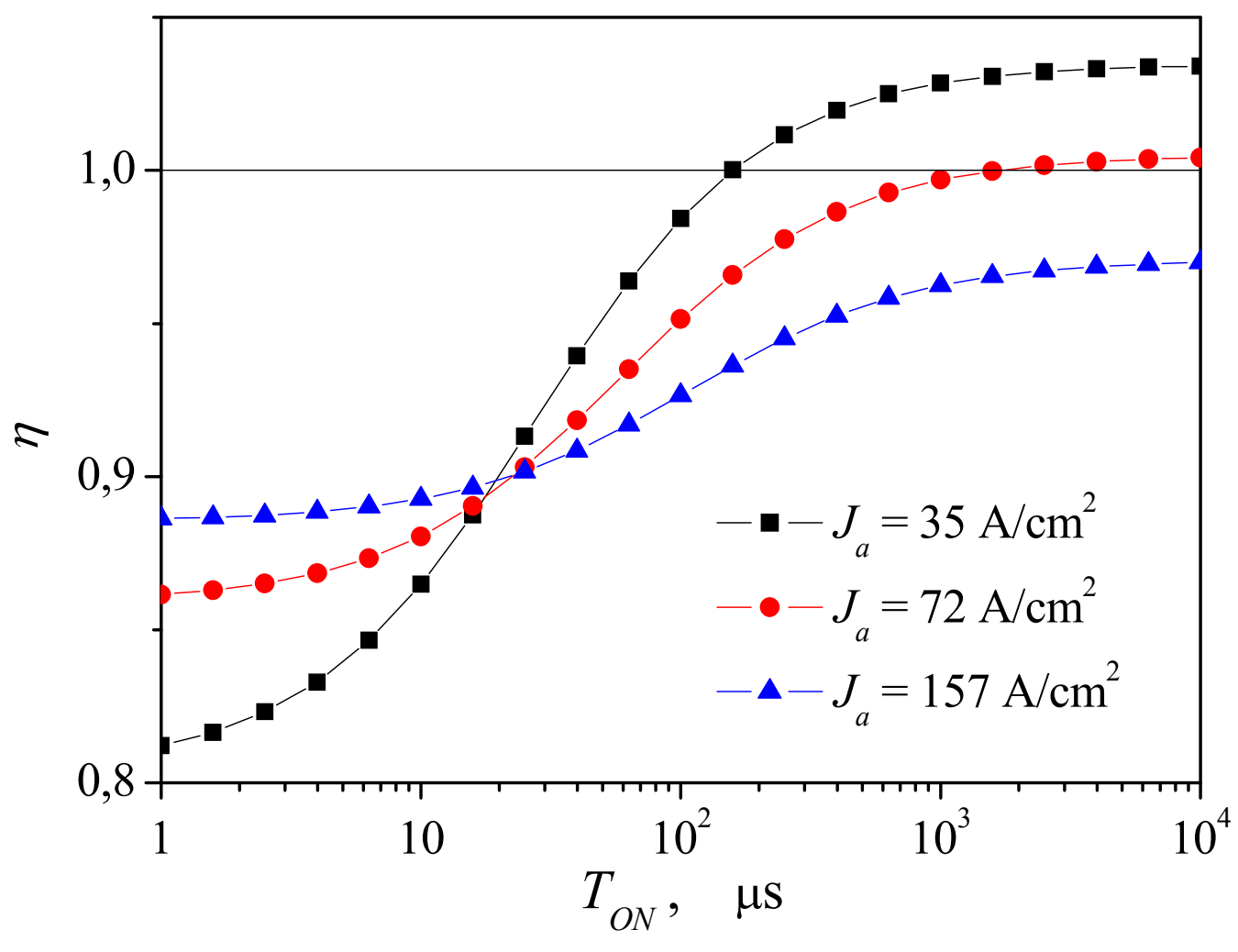

Рис. 5. Зависимость отношения $\eta$ средних мощностей, рассеиваемых КСМТ и CSTBT с $U_b = 1.4$ кВ, от длительности импульса тока с различными амплитудами при $\tau_0 = 383$ нс в CSTBT, $\tau_0 = 300$ нс в КСМТ и $R = 5$ Ом.

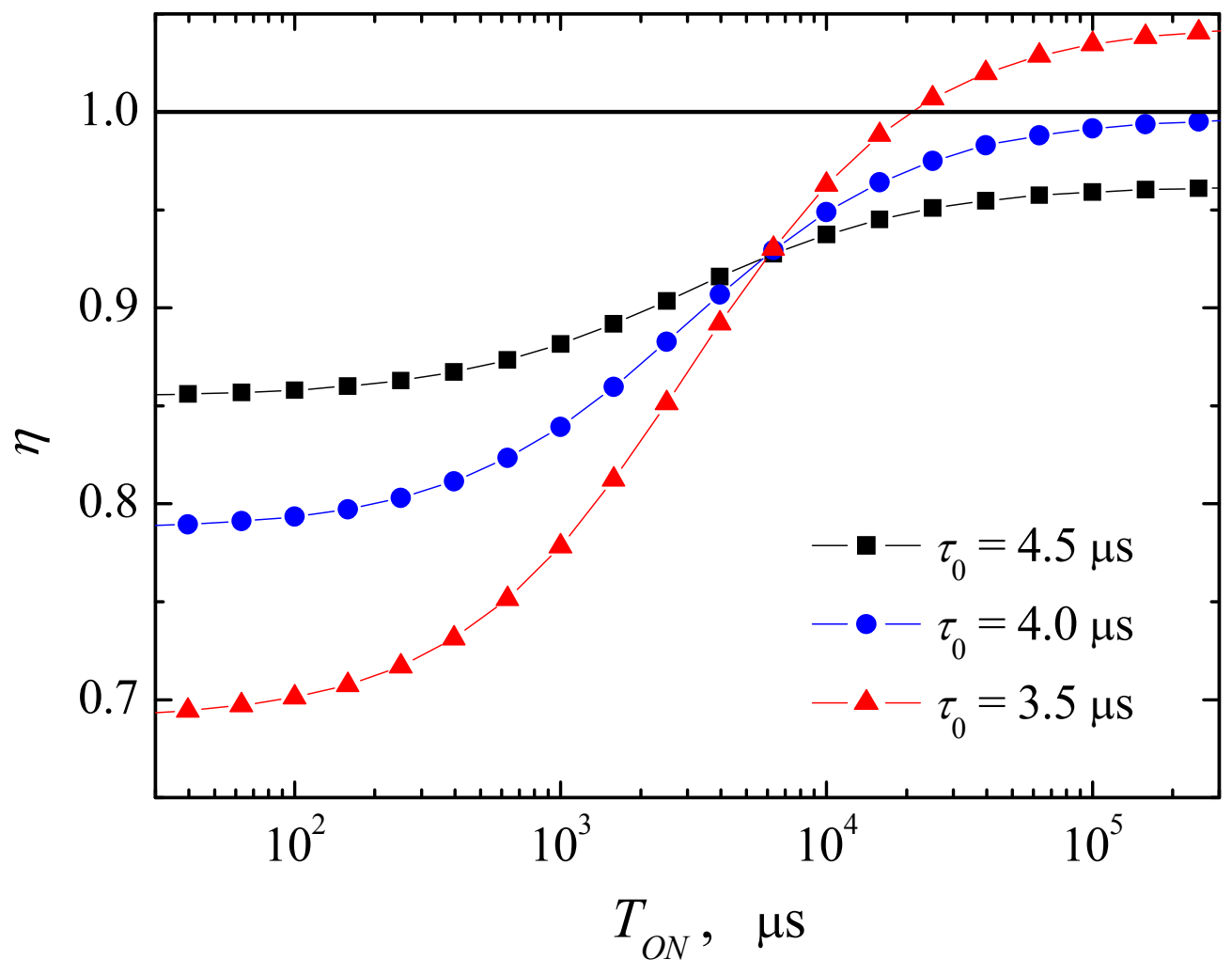

Рис. 6. Зависимость отношения $\eta$ средних мощностей, рассеиваемых КСМТ и CSTBT, от длительности импульса тока с амплитудой 80 А при $U_b = 4.6$ кВ, $R = 30$ Ом, $\tau_0 = 6.13$ мкс в CSTBT и различных $\tau_0$ нс в КСМТ.

Причиной относительно более высоких потерь в CSTBT является низкая эффективность инжекции электронов из катодного эмиттера. Это приводит к сильно неравномерному заполнению высокоомной базы неравновесной плазмой, причем минимум концентрации расположен вблизи катода [1-3,7]. Такое распределение (вариант **А** по терминологии работы [12]) плазмы является худшим из возможных, так как энергия потерь при выключении $W_{off}$ оказывается наибольшей при заданных величинах $Q_{пр}$ или $U_{ON}$ [12,13]. Оптимальным является вариант **С** распределения, при котором минимум концентрации расположен вблизи анода. В этом случае величина $W_{off}$ оказывается в 2-4 раза меньше по сравнению с вариантом **А** при прочих равных условиях [12,13]. Промежуточный вариант **В** с почти равномерным рас-пределением плазмы реализуется в обычных СИТ с постоянной по толщине базы скоростью рекомбинации и близкими коэффициентами инжекции электронов из катода и дырок из анода. В этом случае величина $W_{off}$ оказывается больше, чем в варианте **С**, но меньше, чем в варианте **А**. Именно такой СИТ изучался в настоящей работе, поэтому его преимущество перед CSTBT хотя и несомненное, но не максимально возможное. Очевидно, что уменьшение времени жизни вблизи анода или эффективности инжекции дырок из анода преобразует распределение плазмы в СИТ к оптимальному варианту **С** и еще больше увеличит преимущество КСМТ перед CSTBT. В то же время применение подобных методов к CSTBT в лучшем случае приведет к варианту **В**, но с низкими коэффициентами инжекции и электронов из катода, и дырок из анода.

Работа выполнена при поддержке РФФИ (грант 13-07-00943).

## Литература